# Investigation of New Lead- free (1-x)BaTiO$_3$-xBi(Mg$_{1/2}$Zr$_{1/2}$)O$_3$ Solid Solution with Morphotropic Phase Boundary

Shashwat Anand, Rishikesh Pandey and Akhilesh Kumar Singh*

*Email:akhilesh_bhu@yahoo.com

School of Materials Science & Technology, Indian Institute of Technology (Banaras Hindu University)
Varanasi- 221005,India

## ABSTRACT

We report here the structure and dielectric studies on a new lead free (1-x)BaTiO$_3$-xBi(Mg$_{1/2}$Zr$_{1/2}$)O$_3$ solid solution to explore the morphotropic phase boundary. The powder x-ray diffraction studies on (1-x)BaTiO$_3$-xBi(Mg$_{1/2}$Zr$_{1/2}$)O$_3$ solid solution suggests that structure is tetragonal (P4mm) for the composition with x=0.05 and cubic for the composition with x=0.30 and 0.40. Morphotropic phase boundary is observed in the composition range 0.10<x<0.30, where phase coexistence is observed and composition dependence of room temperature permittivity shows a peak. High temperature dielectric measurement for the composition with x=0.20 exhibits diffuse phase transition having peak temperature around ~ 396 K at 10 kHz. The diffuseness parameter (γ) was obtained to be (1.68± 0.02) for composition with x=0.20.

# INTRODUCTION:

In the last one decade considerable efforts have been put in to explore lead-free alternatives for their toxic lead based counterparts. The Pb based ceramic solid solutions such as $Pb(Zr_xTi_{1-x})O_3$ (PZT), $Pb(Zn_{1/3}Nb_{2/3})O_3$-$PbTiO_3$ (PZN-PT), $(1-x)Pb(Mg_{1/3}Nb_{2/3})O_3$-$xPbTiO_3$ (PMN-PT) etc. have found their way into multiple applications owing to their excellent piezoelectric properties [1]. Other essential features of these Pb-based systems especially PZT include nearly temperature independent morphotropic phase boundary, high Curie temperature (>300 $^0$C for PZT), depolarization temperatures exceeding Curie temperature, poling fields below dielectric breakdown strength etc [2]. Taking cost and toxicity into consideration Na, K and Bi based lead free solid solutions such as $Bi_{0.5}Na_{0.5}TiO_3$ (BNT), $Bi_{0.5}K_{0.5}TiO_3$ (BKT), $KNaNbO_3$ (KNN) were studied extensively as a possible replacement [2]. Other notable examples of lead free piezoelectric ceramics investigated recently are $(K_{0.44}Na_{0.52}Li_{0.04})(Nb_{0.84}Ta_{0.10}Sb_{0.06})O_3$ [3] and $Ba(Ti_{0.8}Zr_{0.2})O_3$-$x(Ba_{0.7}Ca_{0.3})TiO_3$ (BZT-x BCT) [4]

In the past few years a number of $(1-x)BaTiO_3$-$xBiM^{/}O_3$ [BT-$xBM^{/}$] (where, $M^{/}$ is an average trivalent cation $Al^{3+}$ [5], $(Zn_{1/2}Ti_{1/2})^{3+}$ [6], $(Mg_{1/2}Ti_{1/2})^{3+}$ [7,8], $Sc^{3+}$ [9,10], $Fe^{3+}$ [11]) type $BaTiO_3$-based lead free solid solutions have been investigated [5-11]. All these systems show a transition from tetragonal to pseudo cubic phase for increasing concentrations of BM end member. The BT-xBMT system showed this transition clearly for the composition (x=0.04-0.05) [7]. A similar transition was seen in the BT-xBS system for the composition (x=0.05-0.075) [10]. This composition range has also been reported to coexist with tetragonal and rhombohedral phases. Such a variation in symmetry has been shown to exist even in the BT-xBF system before entering the pseudocubic phase at higher compositions [11]. This pseudocubic region continued till higher values of x (for eg. x= 0.075-0.43 in BT-xBS and x=0.07-0.2 for BT-xBMT) before the appearance of a secondary phase.

In the present work, we have synthesized a new lead free solid-solution $(1-x)BaTiO_3$-

$x\text{Bi}(\text{Mg}_{1/2}\text{Zr}_{1/2})\text{O}_3$ ((1-x)BT-xBMZ) aiming to explore the morphotropic phase boundary (MPB) and maximization in electrical properties. $\text{BaTiO}_3$ (BT) is a well- known ferroelectric with the curie temperature $T_C \sim 403$ K (cubic to tetragonal). $\text{Bi}(\text{Mg}_{1/2}\text{Zr}_{1/2})\text{O}_3$ (BMZ) on the other hand has not been synthesized in pure perovskite phase probably due to lower tolerance factor t = 0.922.5. High pressure (5-6 GPa) is required to synthesize the pure perovskite phase of other similar compounds such as $\text{Bi}(\text{Mg}_{1/2}\text{Ti}_{1/2})\text{O}_3$ [12], $\text{Bi}(\text{Zn}_{1/2}\text{Ti}_{1/2})\text{O}_3$ [13] etc. $\text{Bi}(\text{Mg}_{1/2}\text{Zr}_{1/2})\text{O}_3$ ceramic has been stabilized earlier by the formation of its solid solutions with the ferroelectric $\text{PbTiO}_3$ [14]. Here we report its stabilization with $\text{BaTiO}_3$ as the other end member.

## EXPERIMENTAL DETAILS:

The (1-x)BT-xBMZ solid solution was synthesized by conventional solid state ceramics route. Analytical reagent (AR) grade, $\text{Bi}_2\text{O}_3$ (HIMEDIA, 99.5%), $\text{TiO}_2$ (HIMEDIA, 99.0%), $\text{MgCO}_3$ (HIMEDIA, 99.0%), $\text{BaCO}_3$ (HIMEDIA, 99.0%), $\text{ZrO}_2$ (HIMEDIA, 99.0%) were used as raw materials. These powders were weighed at proper molar ratios of (1-x)BT-xBMZ for compositions ranging from x = 0.05 and x = 0.30, and then mixed by ball milling for 6 h. The mixed powders were calcined and re-calcined at an optimized temperature of 990 $^0$C for 5 h. The powders were mixed with PVA (Poly Vinyl Alcohol) as binder. The powders were pressed into pellets using a uniaxial hydraulic press at an optimum load of 60 kN. The binder was burned out at 500 $^0$C for 12 h. The sintering conditions for the pellets were optimized to 1150 $^0$C for 6h. The XRD data was recorded by an 18 Kw rotating Cu anode (Rigaku) powder x-ray diffractometer fitted with monochromator in diffracted beam. The density of the sintered pellets was measured using Archimedes principle. The density of the sintered pellets were higher than >97%. For dielectric measurements, the flat surfaces of the sintered pellets were gently polished with diamond paste and then washed with acetone. Isopropyl alcohol was then applied for removing the moisture. Fired on silver paste was subsequently applied on both the surface of the pellet. It was first dried around 100 $^0$C in an oven and then cured by firing at 500 $^0$C for about 5min. The frequency dependent dielectric constant ($\varepsilon'$) was measured by using Novocontrol (ALPHA A) high performance frequency analyzer. The XRD data were analyzed by Rietveld profile refinement techniques using FULLPROF Suite [15]. In the tetragonal phase with P4mm space group, the asymmetric unit consists of four ions with the Bi/Ba ions in 1(a) sites at (0, 0, z) , Ti/Mg/Zr and $O_I$ in 1(b) sites at (1/2, 1/2, z), and $O_{II}$ in 2(c) sites at (1/2, 0, z). For this space group Bmm2, Bi/Ba ions occupies 2(b) sites at (1/2, 1/2, z), Ti/Mg/Zr ions occupy 2(a) sites at (0, 0, z), $O_I$ in 4(d) sites at (x, 0, z) and $O_{II}$ in the 2(b) sites at (0, 1/2, z). Ti/Mg/Zr ions were fixed at the origin (0, 0, 0) for the refinement. In the cubic phase with space group Pm3m, the occupancy of Bi/Ba were fixed at 1(a) sites (0, 0, 0), Ti/Mg/Zr ions at 1(b) sites (1/2, 1/2, 1/2) and oxygen ions at 3(c) sites (1/2, 1/2, 0).

## RESULTS AND DISCUSSIONS:

Fig.1 shows the XRD profiles of (111), (200) and (220) pseudocubic reflections for the compositions x=0.05, 0.20 and 0.30 of (1-x)BT-xBMZ. It is evident from the Fig.1 that for the composition with x=0.05, (111) pseudocubic reflection is a singlet while (200) pseudocubic reflection shows a doublet with the weaker (002) reflection occurring at lower 2-theta side. Also, the (220) pseudocubic reflections are doublet and show a stronger peak for the (202) reflection at lower 2-theta values. Similar XRD profiles are observed for the room temperature tetragonal phase of $BaTiO_3$ (x = 0.0) [1]. Thus, the structure of the (1-x)BT-xBMZ for the compositions x≤0.05 appears to be tetragonal. The XRD profile of the composition x=0.30 exhibit singlet peaks for all the pseudocubic reflections. The structure of this composition is therefore cubic. All the (111), (200) and (220) XRD profile for the composition x=0.20 show splitting which suggest the phase coexistence for this composition and presence of the MPB. This phase coexistence is a characteristic of the MPB region in ceramic solid-solutions. The composition dependence of permittivity ($\varepsilon'$) at room temperature is shown in Fig.2. It is evident from the Fig.2 that $\varepsilon'$ shows a peak around the composition x=0.18.

To confirm the crystal structures, we performed the Rietveld refinement of the structure taking various plausible space groups such as Bmm2, P4mm and Pm3m. The room temperature (300K) structure of BaTiO$_3$ is tetragonal (P4mm) [1] which transform to the orthorhombic phase (Bmm2) at 5 $^0$C. The Rietveld fit for the composition with x=0.05 using tetragonal with space group P4mm is shown in Fig.3. It is evident from the figure that the fit is quite satisfactory for the space group P4mm, suggest that the structure is tetragonal (P4mm) for this composition. The insets of Fig.3 show the quality of fit for (200) XRD profile. The refined structural parameters with agreement factors is given in Table.1 for the composition with x=0.05.

Very good fit between observed and calculated profiles for the composition with x=0.30 using cubic structure with space group Pm3m is obtained. Rietveld fit for the composition with x=0.30 using cubic structure with space group Pm3m is shown in Fig.4. Very high isotropic thermal parameters (B$_{iso.}$) is obtained for the A-site (B ~ 3.59(3) Å$^2$) and O anions (B ~ 2.04(5) Å$^2$). These high thermal parameters suggest that the system is highly disordered. The A-O (Bi/Ba-O) and B-O (Ti/Mg/Zr-O) bond lengths were 2.85711(4) and 2.02029(4), respectively.

We tried to optimize the stucutre of MPB compositions using various space groups , eg. Pm, Cm, Pm+P4mm, Cm+P4mm, Pm+Cm, P4mm+Pm3m etc. by Rietveld fitting. We got a good fitting with Pm+P4mm and Cm+P4mm phases.  On the other hand Bmm2 + P4mm gives a very bad fit. Detailed study on MPB structure would be given elsewhere.

The most common feature of the BT-xBM$^/$ type systems explored recently is the diffused nature of the tetragonal to cubic phase transition. Moreover, the extent of diffuseness also increases with increasing values of x. The high temperature dielectric study of the system in the present study reveals a similar diffused phase transition. Fig.5(a) shows $\varepsilon^/$ vs T plot for an unpoled sample with the composition x = 0.20. The permittivity peak temperature shows negligible dispersion with frequency, a characteristic of non-relaxor type diffuse phase transition (DPT) found recently in the systems BT-xBF [11] and (1-x)PFN-xPT [16]. The diffuseness of phase transition is described according to the empirical relationship written in equation (i), where $C'$ and $\gamma$ are constants. The parameter $\gamma$ is regarded as a measure of the diffuseness of the phase transition [17-18].

$$1/\varepsilon' - 1/\varepsilon'_{max} = [(T - T'_{max})/C']^\gamma \qquad (i)$$

The value of $\gamma=1$ for normal ferroelectric corresponding to Curie–Weiss law. The complete DPT is observed in relaxors with $\gamma=2$. The non-relaxor type diffuse phase transition generally exhibits intermediate $\gamma$ values between 1 and 2. Fig.5(b) shows the variation of inverse permittivity versus temperature at the frequency 10 kHz for the composition with x=0.20. The curve was fitted with the equation (i) in the temperature range 396 K- 497 K. The upper limit of this range was decided by identifying the region where the curve starts to straighten. The value of $\gamma$ was found to (1.68±0.02) after fitting the inverse of permittivity.

## CONCLUSION:

To summarize, we have proposed a new solid solution in which the structure is tetragonal (P4mm) for the composition with x=0.05 and cubic for the composition with x=0.30. The MPB is observed around x=0.10- 0.25. The composition dependence of the permittivity shows a peak around x=0.18 inside the MPB. The temperature dependence of the $\varepsilon'$ shows a diffuse phase transition ($\gamma$=1.68±0.02) with peak around the $T_C \sim$ 396 K for x=0.20.

FIG1. Selected (111), (200) and (220) XRD profiles of (1-x)BT-xBMZ solid solution sintered at 1150 $^0$C for the compositions with x=0.05, 0.20 and 0.30.

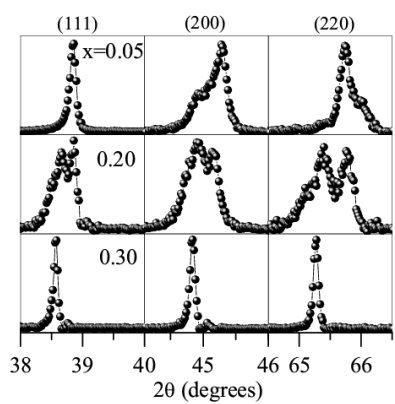

FIG1

FIG2. Room temperature permittivity of (1-x)BT-xBMZ solid solution sintered at 1150 $^0$C for the compositions with x=0.10, 0.15, 0.18, 0.20, 0.22, 0.25 and 0.30 at 10 kHz.

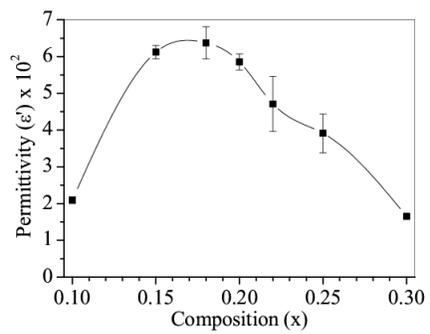

FIG2

3. Observed (dots), calculated (continuous line) and difference (continuous bottom line) profiles for (1-x)BT-xBMZ solid solution with x=0.05 obtained after Rietveld analysis of the powder XRD data using tetragonal (P4mm) structures. The vertical tick marks above the difference plot show the peak positions. The inset figures show the goodness of fitting.

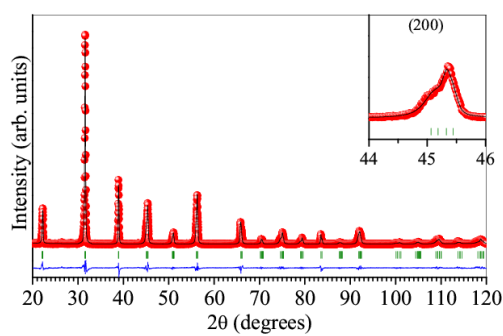

FIG3

FIG4. Observed (dots), calculated (continuous line) and difference (continuous bottom line) profiles for (1-x)BT-xBMZ solid solution with x=0.30 obtained after Rietveld analysis of the powder XRD data using cubic (Pm3m) structures. The vertical tick marks above the difference plot show the peak positions.

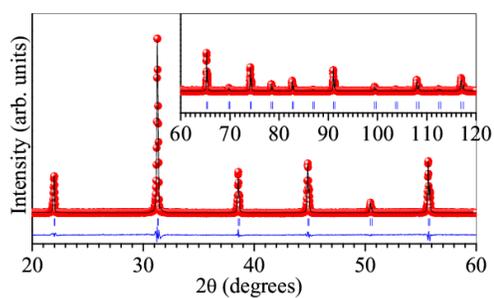

FIG4

FIG5. (a) High temperature permittivity measured in the frequency range 0.5 kHz- 100 kHz (b) Variation of inverse permittivity with temperature fitted in the temperature range 396 K- 497K at 10 kHz, of (1-x)BT-xBMZ solid solutions for the composition with x=0.20.

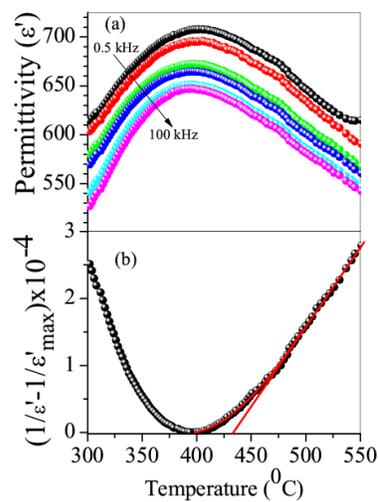

FIG5